\documentclass[journal]{IEEEtran}
\ifCLASSINFOpdf
\else
\fi
\usepackage{amsmath}
\usepackage{tikz}
\usepackage{pgfplots}
\usepackage{pgf}
\usepackage{amssymb}
\usepackage{amsmath}
\usepackage{amsthm}

\usetikzlibrary{external} 
\usepackage[ruled,vlined]{algorithm2e}
\usepackage{cite}

\usepackage[colorinlistoftodos,prependcaption,textsize=tiny]{todonotes}

\definecolor{myblue}{rgb}{0.00000,0.44700,0.74100}
\definecolor{myred}{rgb}{0.85000,0.32500,0.09800}
\definecolor{myyellow}{rgb}{0.92900,0.69400,0.12500}
\definecolor{mypurple}{rgb}{0.49400,0.18400,0.55600}
\definecolor{mygreen}{rgb}{0.46600,0.67400,0.18800}

\begin{document}
%
\title{Successive Eigenvalue Removal\\ for Multi-Soliton Spectral Amplitude Estimation}

\author{Alexander Span, Vahid Aref, Henning B\"ulow, and Stephan ten Brink
\IEEEcompsocitemizethanks{
	\IEEEcompsocthanksitem Date of current version \today. 
	\IEEEcompsocthanksitem A. Span and S. ten Brink are with 
	Institute of Telecommunications, University of Stuttgart, Stuttgart, Germany (E-mail:\{alexander.span,tenbrink\}@inue.uni-stuttgart.de).
	\IEEEcompsocthanksitem V. Aref and H. B\"ulow are with with Nokia Bell Labs, Stuttgart 70435, Germany (E-mail: \{vahid.aref,henning.buelow\}@nokia-bell-labs.com).
	}
}


%


\maketitle

\begin{abstract}
Optical nonlinear Fourier transform-based communication systems require an accurate estimation of a signal's nonlinear spectrum, computed usually by piecewise approximation methods on the signal samples. We propose an algorithm, named successive eigenvalue removal, to improve the spectrum estimation of a multi-soliton pulse. It exploits a property of the Darboux transform that allows removing eigenvalues from the nonlinear spectrum. This results in a smaller pulse duration and smaller bandwidth. The spectral coefficients are estimated successively after removing the eigenvalues of a signal. As a beneficial application, we show that the algorithm decreases the computational complexity by iteratively reducing the pulse duration.

\end{abstract}


\IEEEpeerreviewmaketitle

\section{Introduction}

For the nonlinear optical fiber channel, modeled by the nonlinear Schr\"odinger equation (NLSE), the nonlinear Fourier transform (NFT) maps a pulse from time domain into a so-called nonlinear Fourier domain where the signal transformation along the link can be described by simple equations~\cite{yousefi2014information}. This motivates the modulation of data in the nonlinear spectrum, which needs the computation of the inverse NFT and NFT at the transmitter and receiver, respectively. A special class of transmission pulses, considered here, are multi-solitons which are characterized in the nonlinear spectrum simply by some pairs of eigenvalue and corresponding spectral amplitude. 
In the last few years, multi-soliton communication schemes have been studied and experimentally demonstrated in single polarization transmission, e.g. ~\cite{dong2015nonlinear,aref2015experimental,hari2016multieigenvalue,aref2016onoff,Geisler2016,buelow20167eigenvalues,Gui2017,Koch2019trans}, as well as in polarization-multiplexed transmission, e.g. \cite{Gaiarin2018,Gaiarin2018a,span2019efficient,chan2019experimental}.

 The Darboux transform (DT)~\cite{matveev1991darboux} is a common inverse NFT algorithm to generate multi-soliton signals iteratively.
Although this inverse NFT is rather simple with low complexity,
the forward NFT is a more challenging task in order to compute the eigenvalues and their spectral amplitudes. 
Various numerical methods have been developed so far to compute the spectral amplitudes, see~\cite{yousefi2014information,aref2016control,Hari2016,Vasylchenkova2018,garcia2019statistics,Chimmalgi2018,chimmalgi2019fast,Medvedev20,vaibhav2019efficient,Mullyadzhanov2019}. 
These algorithms are mainly based on numerically solving the Zakharov-Shabat system (ZSS) by a piecewise approximation method. 
The approximation error is rectified usually by a rather large number of signal samples or by using higher-order approximation methods~\cite{Chimmalgi2018,chimmalgi2019fast,Medvedev20,vaibhav2019efficient,Mullyadzhanov2019}.
Higher-order approximation methods increase the computational complexity 
but the error declines faster in terms of signal samples, allowing to find the nonlinear spectrum of a multi-soliton with tens of eigenvalues~\cite{Mullyadzhanov2019}.
Besides, the forward-backward (FB) method (called bi-directional method in \cite{Hari2016}) further reduces
the numerical errors of all one-step discretization NFT methods~\cite{aref2016control}. 
The accuracy of this method is recently improved by some modification in \cite{Prins2019}.

Here, we propose a new approach for the forward NFT based on the Darboux transformation.
Although the DT is commonly known for adding eigenvalues, it is also able to remove eigenvalues from a given solitonic pulse\cite{Junmin1990}. Applying the DT carefully removes an eigenvalue from the pulse's nonlinear spectrum with a simple change on the remaining nonlinear spectrum of the pulse. As we explain in Sec.~\ref{sec:basic}, $a(\lambda)$
will be changed but the $b-$values remain the same.
While this property of the DT is known for a long time, it has not yet been used for the detection of the nonlinear spectrum.

We propose the Successive Eigenvalue Removal (SER) algorithm where the spectral coefficients are estimated one by one successively. Following a predefined order, an eigenvalue and its spectral coefficient are estimated and then removed from the nonlinear spectrum. 
This process is repeated on the remaining pulse until all eigenvalues are removed.
This algorithm is beneficial as removing an eigenvalue can reduce the pulse duration as well as the bandwidth (smoother variations). Both reductions can decrease the number of numerically required pulse samples and thus decrease the total computational complexity of the NFT computation. 

We discuss that the SER algorithm is quite robust against the error propagation drawback of successive algorithms in which usually an error in an early iteration influences the result of the following iterations and thus the error accumulates.   
 We compare the SER algorithm to the conventional approach in terms of computational complexity and numerical accuracy by simulation, also in the presence of noise.
Note that the SER algorithm can be equipped with any aforementioned forward NFT algorithms and can improve the performance of those algorithms. 
If the continuous spectrum contains a large fraction of the pulse energy (not the case for multi-solitons), the SER algorithm can still be helpful in computing the discrete spectrum by simplifying the pulse after an eigenvalue removal, but it may not lead to a reduction in pulse duration.

The paper is outlined as follows: In Sec.~\ref{sec:basic}, we review the basics of the NFT 
and describe how the DT can remove eigenvalues from a nonlinear spectrum.
We explain the SER algorithm in Sec.~\ref{sec:SolDecom} and explain its robustness against coarse estimation of eigenvalues in Sec.~\ref{sec:analysis}. We compare its performance numerically to the conventional approaches in Sec.~\ref{sec:Sim_eval}. We draw some conclusions in Sec.~\ref{sec:concl}.

\section{Preliminaries of NFT and Darboux Transform}\label{sec:basic}
In a fiber link with ideal distributed amplification, the pulse propagation in a single polarization can be modeled by the nonlinear Schr\"odinger equation (NLSE),
\begin{equation}\label{eq:NLSE}
j\frac{\partial q(t,z)}{\partial z}=\frac{\partial^2 q(t,z)}{\partial t^2}+2\left\|q(t,z)\right\|^2 q(t,z).	
\end{equation}
where $t$, $z$ and $q(t,z)$ are the normalized time, distance along the fiber and the normalized pulse envelope, respectively. On a single mode fiber with chromatic dispersion $\beta_2$ and Kerr nonlinearity $\gamma$, the physical pulse is obtained simply from
\begin{equation*}
p(\tau,\ell)=\sqrt{P_0}q\left(\frac{\tau}{T_0},\ell\frac{|\beta_2|}{2T_0^2}\right), \quad  \text{with } P_0T_0^2=\frac{|\beta_2|}{\gamma}
\end{equation*}
where $T_0$ is an arbitrary time-scale, $\tau$ and $\ell$ are the physical time and distance along the fiber.

\subsection{Basics of Nonlinear Fourier Transform}\label{sec:NFT}

A solution of \eqref{eq:NLSE} can be represented in a nonlinear spectrum via the Zakharov Shabat system (ZSS) \cite{shabat1972exact},
\begin{equation}\label{eq:ZS}
\frac{\partial}{\partial t}\left( \begin{matrix} \vartheta_1(\lambda;t,z) \\ \vartheta_2(\lambda;t,z) \end{matrix} \right)
= \left( \begin{matrix} -j\lambda & q(t,z) \\ -q^*(t,z) & j\lambda \end{matrix} \right)
\left( \begin{matrix} \vartheta_1(\lambda;t,z) \\ \vartheta_2(\lambda;t,z) \end{matrix} \right)
\end{equation}
with the boundary condition
\begin{equation}
\left( \begin{matrix} \vartheta_1(\lambda;t,z) \\ \vartheta_2(\lambda;t,z) \end{matrix} \right)\to \left( \begin{matrix} 1 \\ 0 \end{matrix} \right) \exp(-j\lambda t)
\qquad \mathrm{for} \quad t \to -\infty	\label{eq:boundary_condition_left}
\end{equation}
under the vanishing boundary assumption $q(t)\to 0$ as $t \to \pm \infty$.
The nonlinear (Jost) coefficients are defined as 
\begin{align}
a(\lambda;z)&=\lim_{t\to +\infty} \vartheta_1(\lambda;t,z) \exp(j\lambda t)	\nonumber
\\
b(\lambda;z)&=\lim_{t\to +\infty} \vartheta_2(\lambda;t,z) \exp(-j\lambda t). \label{eq:spec_coeffs}
\end{align}
The crucial property of the nonlinear spectrum is
\begin{equation}\label{eq:evolution_z}
a(\lambda;z)=a(\lambda;0), \quad b(\lambda;z)=\exp(-4j\lambda^2 z)b(\lambda;0).
\end{equation}
This property indicates that the nonlinear spectrum for each $\lambda$
evolves independently in $z$ according to this simple equation. This suggests to modulate information on the nonlinear frequencies.
We drop the dependency on $z$, as it is simply multiplicative in \eqref{eq:evolution_z}.

The nonlinear spectrum of $q(t)$ consists of two parts:
\begin{itemize}
\item[(i)] the \emph{discrete spectrum} $\{(\lambda_k,b_k)\}$, where
the $N$ eigenvalues ${\lambda_k=\omega_k+j\sigma_k \in \mathbb{C}^+}$ ($\omega_k,\sigma_k \in \mathbb{R}$) are the roots of $a(\lambda)$ and spectral coefficients $b_k=b(\lambda_k)$.

\item[(ii)] the \emph{continuous spectrum} $Q_c(\lambda)=\frac{b(\lambda)}{a(\lambda)}$ for ${\lambda \in \mathbb{R}}$.
\end{itemize}

Note that the discrete spectrum is sometimes defined differently 
like in \cite{yousefi2014information}. The above representation is chosen as $b_k$ are more preferable for data modulation~\cite{Aref2018,Gui2017,Buelow2018,yangzhang2019dual}.

There are various ways to compute the nonlinear spectrum by numerically solving \eqref{eq:ZS} \cite{yousefi2014information,wahls2015fast,Turitsyn2017,Chimmalgi2018,Medvedev20}. Practically, it is assumed that the signal $q(t)$ is truncated such that it is non-zero only inside the interval $t \in [T_-,T_+]$. Then the differential equation system \eqref{eq:ZS} can be solved by propagating $\boldsymbol{\vartheta}(\lambda;t)$ from the known solution \eqref{eq:boundary_condition_left} at the boundary $t=T_-$ to $t=T_+$. The time is discretized with step size $h$ as $t_m=T_- + m\cdot h$ where the sample index is $m=0,1,\dots,M-1$.

By applying a change of variables  
\begin{equation}\label{eq:var_change}
\boldsymbol{\Psi}(\lambda;t)=\left( \begin{matrix} \vartheta_1(\lambda;t) \exp(j\lambda t) \\ \vartheta_2(\lambda;t) \exp(-j\lambda t)
\end{matrix} \right),
\end{equation}
and defining $\boldsymbol{w}^{(m)}=\boldsymbol{\Psi}(\lambda;t_m)$, one can iterate
\begin{equation}\label{eq:vec_update}
\boldsymbol{w}^{(m+1)} = \boldsymbol{S}_{m} \boldsymbol{w}^{(m)}
\end{equation}
from $\boldsymbol{w^{(0)}}=\boldsymbol{\Psi}(\lambda;T_-)=\left( 1,0 \right)^T$ (obtained from the boundary condition \eqref{eq:boundary_condition_left}) to $\boldsymbol{w^{(M)}}\approx \boldsymbol{\Psi}(\lambda;T_+)$. Then, according to \eqref{eq:spec_coeffs},\eqref{eq:var_change}, the spectral coefficients are obtained from
\begin{align}\label{eq:spec_coeffs_estimation}
\hat{a}(\lambda)=w^{(M)}_1,
\qquad
\hat{b}(\lambda)=w^{(M)}_2.
\end{align}
The scattering matrix $\boldsymbol{S}_{m}$ depends on the approximation method.
Using mid-point approximation, it becomes
\begin{equation}\label{eq:iteration_matrix}
\boldsymbol{S}_{m} = \left( \begin{matrix} \cos(|q_m h|) & \frac{q_m \sin(|q_m| h) \exp(2j\lambda t_m)}{|q_m|}  \\ -\frac{q_m^* \sin(|q_m| h) \exp(-2j\lambda t_m)}{|q_m|} & \cos(|q_m| h) \end{matrix} \right)
\end{equation}
where $q_m$ are the samples $q(t_m)$ at time $t_m$. Applying a higher-order approximation method, e.g. \cite{Chimmalgi2018,Medvedev20}, results in different $\boldsymbol{S}_{m}$ with smaller numerical errors.

The forward-backward method \cite{aref2016control} is a modification of the iteration \eqref{eq:vec_update}. In addition to propagating $\boldsymbol{w}^{(m)}$ from $t=T_-$ to $t=T_+$, it propagates another vector $\boldsymbol{u}^{(m)}$ from $t=T_+$ to $t=T_-$ with boundary condition
$\boldsymbol{u}^{(M)}=\left( 0,1 \right)^T$, i.e.
\begin{equation}\label{eq:vec_update_FB}
\boldsymbol{u}^{(m-1)} = \boldsymbol{S}_{m}^{-1} \boldsymbol{u}^{(m)}
\end{equation}
where $\boldsymbol{S}_{m}^{-1}$ is the inverse of $\boldsymbol{S}_{m}$, obtained by changing $h$ to $-h$ in $\boldsymbol{S}_{m}$.
Both iterations \eqref{eq:vec_update} and \eqref{eq:vec_update_FB} are repeated up to some index $0<p<M$. For $\lambda=\lambda_k$ being an eigenvalue,
the spectral coefficients are then obtained as
\begin{align}\label{eq:spec_coeffs_estimation_FB}
& \hat{a}(\lambda_k)=w_1^{(p)} u_2^{(p)} - u_1^{(p)} w_2^{(p)}\approx 0	\nonumber
\\
& \hat{b}(\lambda_k)=\frac{w_2^{(p)}}{u_2^{(p)}}
\end{align}

In the next section, we use the following approximation of $\boldsymbol{\vartheta}(\lambda_k;t)=\left(\vartheta_1(\lambda_k;t),\vartheta_2(\lambda_k;t) \right)^\mathrm{T}$. For $m<p$,
\begin{equation}
\hat{\boldsymbol{\vartheta}}(\lambda_k;t_m) \approx \left( \begin{matrix} w_1^{(m)} \exp(-j \lambda_k t_m) \\ w_2^{(m)} \exp(j \lambda_k t_m) \end{matrix} \right)
\end{equation}
while for $m>p$ 
\begin{equation}
\hat{\boldsymbol{\vartheta}}(\lambda_k;t_m) \approx \hat{b}(\lambda_k) \left( \begin{matrix} u_1^{(m)} \exp(-j\lambda_k t_m) \\ u_2^{(m)} \exp(j\lambda_k t_m) \end{matrix} \right)  
\end{equation}
We have observed that $p=\frac{M}{2}$ gave rather precise approximations 
for the pulses in the next sections. In general, one can choose $p$ according to the criterion in \cite{Prins2019}.

\subsection{Two Properties of the Darboux Transform}\label{sec:DT}

Multi-soliton pulses are specific solutions of the NLSE having only a discrete spectrum. 
To numerically generate a multi-soliton pulse, an efficient algorithm is based on the DT. 
It constructs a signal by adding eigenvalues one by one recursively and updating the pulse accordingly.
The DT is an elegant approach to modify a pulse by adding a new eigenvalue or also removing an existing one while the rest of the discrete spectrum as well as $b(\lambda)$, $\lambda \in \mathbb{R}$, are unchanged~\cite{Junmin1990}.
Let us briefly summarize the DT and its properties.
Assume a pulse $q^{(n)}(t)$ with $n$ eigenvalues $\Lambda^{(n)}=\{\lambda_1,\dots,\lambda_n\}$.
Consider a complex frequency $\mu\in\mathbb{C}^+$. Let $\boldsymbol{\vartheta}^{(n)}(\mu;t)$ denote a solution of \eqref{eq:ZS} for the pulse $q^{(n)}(t)$. By applying the following transformation,
\begin{equation}\label{eq:sig_update}
\tilde{q}(t) = q^{(n)}(t) +\frac{2j(\mu^*-\mu) \vartheta_{2}^{*(n)}(\mu;t)\vartheta_{1}^{(n)}(\mu;t)}{|\vartheta_1^{(n)}(\mu;t)|^2+|\vartheta_2^{(n)}(\mu;t)|^2}
\end{equation}
the spectrum of $\tilde{q}(t)$ is related to the one of $q^{(n)}(t)$ according to the following two cases:

\subsubsection{Adding Eigenvalues} If $\mu\not\in \Lambda^{(n)}$, then $\tilde{q}(t)$ has $n+1$ eigenvalues $\Lambda^{(n+1)}=\Lambda^{(n)}\cup\{\mu\}$. In this case, we write $\tilde{q}(t)=q^{(n+1)}(t)$. 
As shown in \cite{aref2016control,Aref2018,Junmin1990}, the required condition to enforce
the desired $b_{n+1}$ is
\begin{equation}
    \lim_{t\to-\infty} e^{-j\mu t}\vartheta_2^{(n)}(\mu;t) = -b_{n+1}\lim_{t\to+\infty} e^{+j\mu t}\vartheta_1^{(n)}(\mu;t).
\end{equation}

Interestingly, $b(\lambda)$ for $\lambda\in\mathbb{R}$ and $b_k$ for $\lambda_k\in\Lambda^{(n)}$ of $q^{(n+1)}(t)$ stay the same as the ones of $q^{(n)}(t)$~\cite{Junmin1990}. This property is commonly used to generate a multi-soliton recursively~\cite{matveev1991darboux}, or to generate a pulse with a continuous and discrete spectrum~\cite{Aref2018}.

\subsubsection{Removing Eigenvalues} 
\begin{figure}[tb]
\centering
\includegraphics{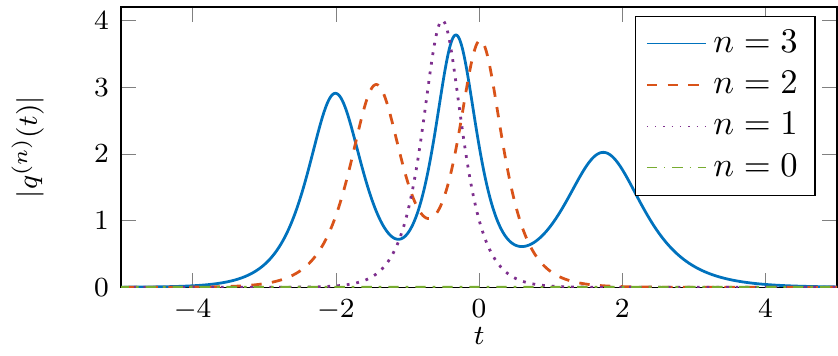}
%
%
%
%
%
%
%
%

\caption{Example of third order soliton decomposition with initial discrete spectrum $(\lambda_k,b_k)\in\{(1j,2.1),(1.5j,-0.09),(2j,0.13)\}$. The dashed pulse is resulted after removal of $\lambda_1=1j$ and the dotted pulse is resulted after removal of $\lambda_1=1j$ and $\lambda_2=1.5j$.
}
\label{fig:example_decomposition}
\end{figure}

If $\mu\in \Lambda^{(n)}$, then $\tilde{q}(t)$ has $n-1$ eigenvalues $\Lambda^{(n-1)}=\Lambda^{(n)}\setminus\{\mu\}$. In this case, we write $\tilde{q}(t)=q^{(n-1)}(t)$. 
Without loss of generality, assume that $\mu=\lambda_n$. Then the relation between the nonlinear spectrum of $q^{(n-1)}(t)$
and the one of $q^{(n)}(t)$ is given by,

\begin{align}
& a^{(n-1)}(\lambda)=\frac{\lambda-\lambda_n^*}{\lambda-\lambda_n}a^{(n)}(\lambda), \text{ for } \lambda\in\mathbb{C}\\
& \Lambda^{(n-1)}=\{\lambda_1,...,\lambda_{n-1}\} \\
& b^{(n-1)}(\lambda)=b^{(n)}(\lambda), \text{ for } \lambda\in\mathbb{R} \\
& b^{(n-1)}_k=b^{(n)}_k, \text{ for eigenvalue } \lambda_k,1\leq k\leq n-1. 
\end{align}
Consequently, the remaining eigenvalues and corresponding spectral coefficients do not change if the signal update \eqref{eq:sig_update} is applied. In Fig.~\ref{fig:example_decomposition} we show an example of successively applying \eqref{eq:sig_update} on a multi-soliton pulse. The pulse has initially 3 eigenvalues which are removed successively in ascending order of their imaginary part.
After the last eigenvalue removal, no residual pulse is left. 

\section{Successive Eigenvalue Removal Algorithm}\label{sec:SolDecom}

\subsection{Motivation}

Finding the discrete spectrum from \eqref{eq:ZS} is not an easy task. In practice, the tails of a pulse are truncated and then
\eqref{eq:ZS} is approximated from the pulse samples. These two steps cause some approximation errors in finding the eigenvalues and spectral amplitudes. A discretized NFT like \eqref{eq:vec_update} accumulates the approximation errors. This error is very sensitive to the sampling rate and the truncation. When a pulse has a large support, usually a large sampling rate is required to achieve small errors. A typical observation is that a larger number of eigenvalues in a pulse results in either faster variations of the pulse shape or a longer pulse duration or both. 
Both effects increase the approximation errors. 
In contrast, removing an eigenvalue may lead to less variations and a reduction in pulse duration.
Both effects decrease the number of required samples which leads to a lower computational complexity.

Let $q^{(n)}(t)$ denote a multi-soliton pulse with $n$ eigenvalues $\lambda_k=\omega_k+j\sigma_k$, $k=1,\dots,n$.
Assume that the eigenvalues are indexed in decreasing order of their imaginary part, i.e. 
$\sigma_1>\sigma_{2}> \dots> \sigma_n$. It is shown in \cite{Span2019} that when $t\to\pm\infty$
\begin{align}
\label{eq:tail}
|q^{(n)}(t)| \sim & 4\sigma_n |b_n|^{\pm 1} \exp(\mp 2\sigma_n(t \mp t_s)),
\\
\mathrm{with} \qquad & t_s=\frac{1}{2\sigma_n}\ln \left( \left| \prod_{k=1}^{n-1} \frac{\lambda_n - \lambda_k^*}{\lambda_n - \lambda_k} \right| \right)
\label{eq:ts}.
\end{align}
When $|b_k|\sim 1$ for all eigenvalues (i.e. they are in the same order) and eigenvalues are not located
too close together (as in practice),
removing $\lambda_n$ (having the smallest imaginary part) decreases $t_s$ and also 
increases the exponent of \eqref{eq:tail} resulting a faster decay in the tails.
Both changes can be seen in Fig.~\ref{fig:example_decomposition}.
Note that removing other eigenvalues reduces also the pulse duration slightly as $t_s$ decreases and the DT enables us to remove eigenvalues in any desired order.

\begin{figure}[tb]
\center
\includegraphics[scale=0.65]{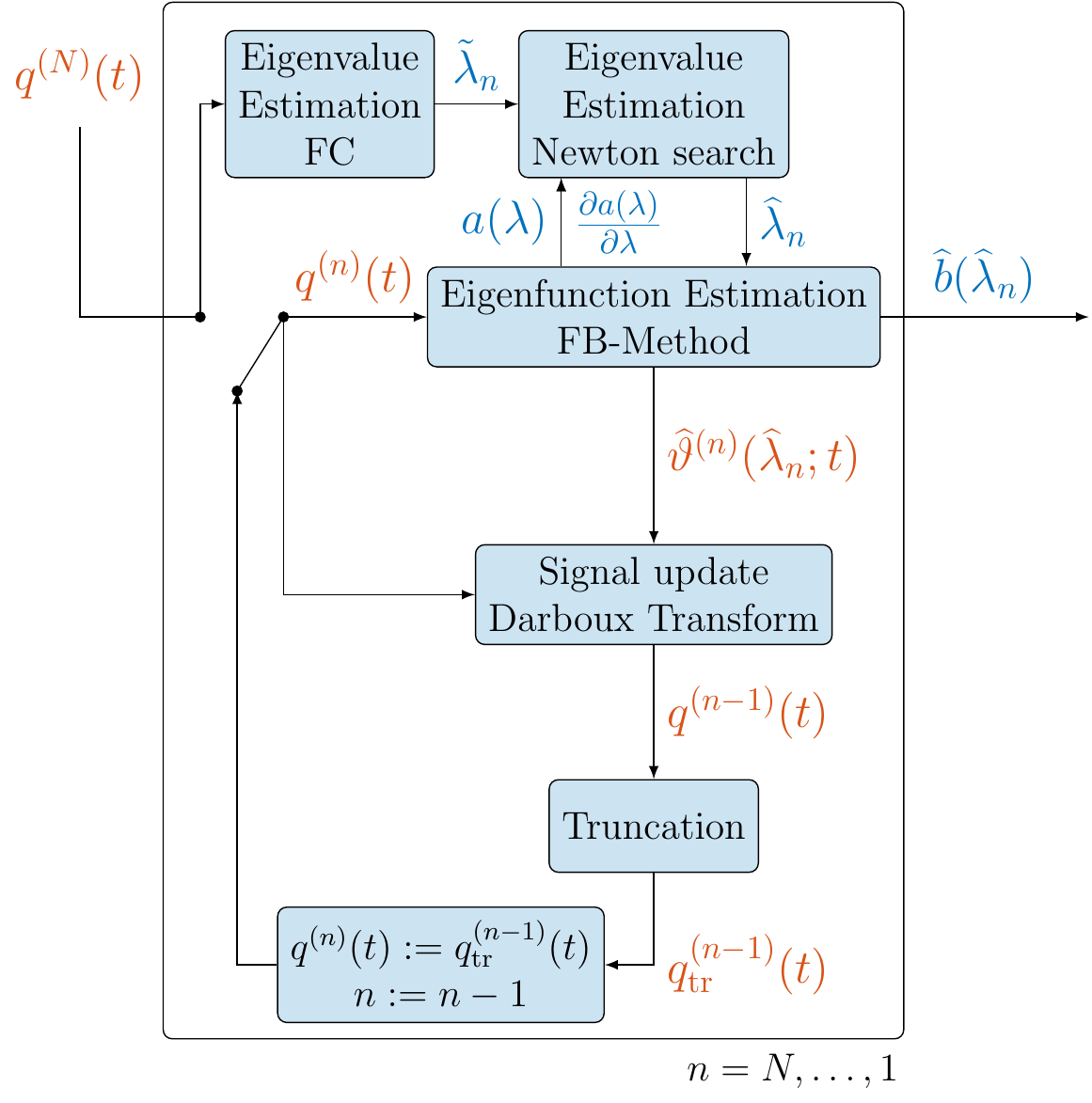}
\caption{Block diagram of the Successive Eigenvalue Removal algorithm.}
\label{fig:algorithm}
\end{figure}

\subsection{Eigenvalue Removal and Truncation}

Let $q^{(n)}(t)$ denote a multi-soliton pulse
as defined in the previous section.
We define the effective pulse support as the smallest time interval $[T_-,T_+]$
such that $|q^{(n)}(t)|\leq 2\sigma_n\sqrt{\varepsilon}$ for $t \notin [T_-,T_+]$ and some small fixed $\varepsilon$.
The pulse tails outside $[T_-,T_+]$ are truncated and the pulse duration is then $T=T_+-T_-$. While other definitions for pulse duration can be taken as well, this definition has the advantage that 
the truncation threshold scales the same as the pulse energy by normalization of the NLSE \eqref{eq:NLSE}.
For instance, the effective support of a first-order soliton with eigenvalue $j\sigma_n$ contains $\sqrt{1-\varepsilon}$ fraction of the pulse energy regardless of  $\sigma_n$ value. 

As it is shown in Fig.~\ref{fig:algorithm}, the Successive Eigenvalue Removal (SER) algorithm computes the discrete spectrum of a multi-soliton $q^{(N)}(t)$ in few steps.
First, it requires an initial guess of eigenvalues 
$\tilde{\lambda}_k$, $k=1,\dots ,N$ to determine the sequential order of removal
and the effective pulse support after each iteration.
We use the Fourier collocation (FC) method~\cite{yang2010nonlinear,garcia2019statistics} but one can use any other estimator, e.g. the one finding $\tilde{\lambda}_k$ from the continuous spectrum~\cite{aref2019efficient}.
This step can be skipped when an initial guess of eigenvalues is available.
Here, we remove the eigenvalues in the increasing order of their imaginary part as follows: 

\begin{itemize}
    \item[(i)] [\emph{Refinement of eigenvalue}]
    Consider the eigenvalue with smallest imaginary part $\tilde{\lambda}_n$.
    If it is necessary, we improve the eigenvalue estimation further 
    via a Newton zero search $a(\lambda)=0$ in the neighborhood of $\tilde{\lambda}_n$.
    For that, we estimate $a(\lambda)$ and $\frac{\partial a(\lambda)}{\partial \lambda}$ according to Sec.~\ref{sec:NFT} based on the current pulse $q^{(n)}(t)$. We end up with a more precise eigenvalue estimate $\hat{\lambda}_n$.  
    
    \item[(ii)] [\emph{Spectral amplitude estimation}] We numerically compute the solution $\hat{\boldsymbol{\vartheta}}^{(n)}(\hat{\lambda}_n;t)$ for $t \in [T_-^{(n)},T_+^{(n)}]$ as described in Sec.\ref{sec:NFT}. Then, we obtain $\hat{b}(\hat{\lambda}_n)$ from \eqref{eq:spec_coeffs_estimation_FB}.
    
    \item[(iii)] [\emph{Eigenvalue removal}] We apply the Darboux update \eqref{eq:sig_update} to $q^{(n)}(t)$ using $\hat{\boldsymbol{\vartheta}}^{(n)}(\hat{\lambda}_n;t)$ for removing the eigenvalue $\lambda_n$ (the eigenvalue with smallest imaginary part) from the nonlinear spectrum. The resulting pulse is $q^{(n-1)}(t)$ and has $n-1$ eigenvalues. 
    
    \item[(iv)] [\emph{Pulse truncation}] The pulse support is truncated to the interval $[T_-^{(n-1)},T_+^{(n-1)}]$ with the truncation threshold of $2\mathrm{Im}\{\hat{\lambda}_{n-1}\}\sqrt{\varepsilon}$. 

\end{itemize}

We repeat these steps until all $N$ eigenvalues are removed. 
Note that we assume a very small $\varepsilon$ to avoid a severe
pulse truncation causing changes in eigenvalues as well as $b-$values. We refer the interested readers to \cite{vaibhav2018nonlinear,aref2019nonlinear,gelash2019direct} for the spectral changes after truncation. In the case of severe truncation, the SER algorithm is still applicable and 
will find the distorted spectrum (like any other NFT algorithm).

It is worth mentioning that the SER algorithm works in principle 
for any removal order of eigenvalues. Here, we remove eigenvalues in
increasing order of their imaginary part. This is a suitable choice when $\frac{1}{\sigma_k}\ln(|b_k|)$
have almost the same magnitudes but it may not 
be the best choice in other cases to improve complexity or accuracy.
For instance, when a pulse consists of two separate solitonic compenents in time-domain ($\frac{1}{\sigma_k}\ln(|b_k|)$ have large variance), then 
it may be more efficient to remove first the eigenvalues of a solitonic component and then remove the ones of the other component.

\begin{figure}[tb]
\centering
\includegraphics{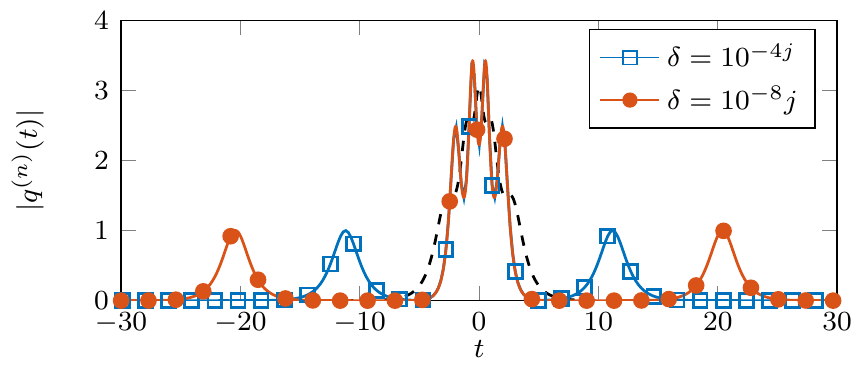}
\includegraphics{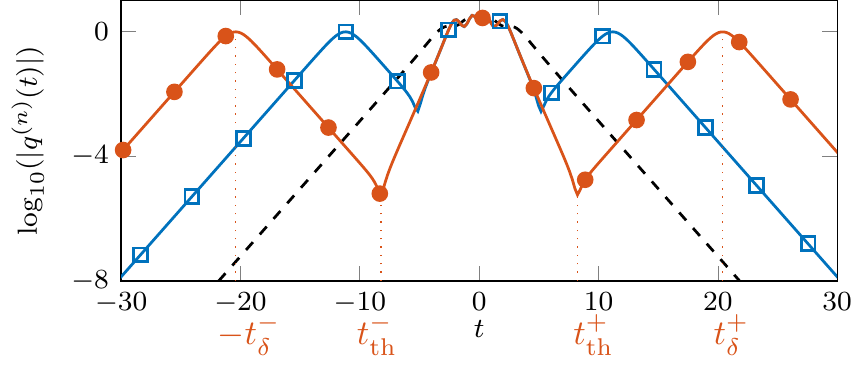}

\caption{Addition of $(\hat{\lambda}_n=0.5j+\delta,\hat{b}_n=\exp(5.88j))$ 
to a multi-soliton
(dashed line) with five eigenvalues $\Lambda^{(5)}=\{2.5j,2j,1.5j,1j,0.5j\}$, and $b_k=\exp(j\varphi_k)$ of randomly chosen $\varphi_k=0.24,4.90,0.58,3.98,0.09$, respectively. The modified pulses $\tilde{q}(t)$ (solid lines) have three parts: two first-order soliton with eigenvalue $\approx 0.5j$ and a multi-soliton with four eigenvalues
$\Lambda^{(4)}=\{2.5j,2j,1.5j,1j\}$ in the middle. The distance of two first-order solitons is $t^+_\delta+t^-_\delta=O(-\ln(|\delta|))$, given in \eqref{eq:t_delta}. Both figures show the same plots with different y-axis in linear (top) and logarithmic (bottom) scale. 
}
\label{fig:6Sol_closeEV}
\end{figure}

\section{Analysis of Eigenvalue Estimation Error}\label{sec:analysis}

An existing eigenvalue $\lambda_n$ will be removed by DT from the discrete spectrum, if $\boldsymbol{\vartheta}^{(n)}(\mu;t)$ is computed exactly at $\mu=\lambda_n$. In this section, 
we explain how the eigenvalue estimation error $\delta = \hat{\lambda}_n-\lambda_n$ affects the capability of the SER algorithm to remove the eigenvalue $\lambda_n$.
As before, we assume that $\lambda_n$ is the eigenvalue with the smallest imaginary part.

Since there is always an estimation error $\delta \neq 0$, the SER algorithm indeed adds a new eigenvalue $\hat{\lambda}_n$ to the spectrum.
The SER applies the DT signal update \eqref{eq:sig_update} at $\mu=\hat{\lambda}_n$ where $\hat{\mathbf{\vartheta}}^{(n)}(\mu;t)$ is numerically computed from the pulse $q^{(n)}(t)$.
Then, the resulting pulse $\tilde{q}(t)$ has an additional eigenvalue $\hat{\lambda}_n=\lambda_n+\delta$. 
Note that there are two kinds of error: $(i)$ estimation error $\delta$ $(ii)$ numerical errors in $\hat{\mathbf{\vartheta}}^{(n)}(\hat{\lambda}_n;t)$. To study only the effect of $\delta$, we use the DT to generate a multi-soliton
with discrete spectrum $\{(\lambda_k,b_k)\}_{k=1}^n$ and additionally $\{(\hat{\lambda}_n,\hat{b}_n)\}$ for some arbitrary value $\hat{b}_n$.
When adding $\hat{\lambda}_n$ in the last iteration, one can generate 
the exact $\mathbf{\vartheta}^{(n)}(\hat{\lambda}_n;t)$ as well as $\tilde{q}(t)$. The detailed algorithm is given in \cite{garcia2019statistics} 
and we skip it here.

When $|\delta|$ is small enough, adding $\hat{\lambda}_n$ decomposes $\tilde{q}(t)$ into three separate parts in time domain: two first-order solitons with eigenvalues $\hat{\lambda}_n\approx \lambda_n$ and another pulse between them containing the rest of the discrete spectrum.
The separation effect is illustrated in Fig.~\ref{fig:6Sol_closeEV}.
A multi-soliton with 
five eigenvalues $\Lambda^{(5)}=\{2.5j,2j,1.5j,1j,0.5j\}$ (dashed line)
is modified by an additional eigenvalue $\hat{\lambda}=0.5j+\delta$ for
two choices of $\delta$. One can observe that 
the first-order solitons (sech-shape pulses in the tails) are pushed away from the middle part as $|\delta|$ decays.
We can analytically approximate the distance between two first-order solitons in terms of $\delta$
and spectral amplitude $\hat{b}_n$. Assume that $|\delta|\ll \sigma_n$ and denote $b_n=|b_n|\exp(j\varphi_n)$ and $\hat{b}_n=|\hat{b}_n|\exp(j\hat{\varphi}_n)$.
Similar to the analysis in \cite{Span2019}, we write 
the first-order approximation of the DT update \eqref{eq:sig_update} to characterize
$\tilde{q}(t)$ when $t\to\pm\infty$,
\begin{align}
\tilde{q}(t) \sim &
-2\sigma_{n} e^{\mp j \varphi_{\delta}^{\pm} - 2\omega_n t}\mathrm{sech}\left( 2\sigma_{n}(t\mp t_{\delta}^{\pm})\right) \label{eq:qtilde_tails}
\\
\mathrm{with} \qquad & t_{\delta}^{\pm} \approx \frac{1}{2 \sigma_n} \ln \left( \left| \frac{2j\sigma_n}{\delta} \prod_{k=1}^{n-1} \frac{\lambda_n-\lambda_k^*}{\lambda_n-\lambda_k} \alpha^\pm \right| \right)	\label{eq:t_delta}
\\
& \varphi_{\delta}^{\pm} \approx \arg \left( \frac{2j\sigma_n}{\delta} \prod_{k=1}^{n-1} \frac{\lambda_n-\lambda_k^*}{\lambda_n-\lambda_k} \alpha^\pm \right)
\\
& \alpha^\pm= \left(|\hat{b}_n|^{\pm 1} e^{\pm j\hat{\varphi}_n} - |b_n|^{\pm 1} e^{\pm j\varphi_n}  \right)
\end{align}
We found in different simulations that the above $t_{\delta}^{+}$ and $t_{\delta}^{-}$ are very close to their numerical values. These values are shown in Fig.~\ref{fig:6Sol_closeEV} for $\delta=10^{-8}j$. 

The distance between two first-order solitons is $t_{\delta}^{+} + t_{\delta}^{-}=O(\ln(\frac{1}{|\delta|}))$. Look again at Fig.~\ref{fig:6Sol_closeEV}. Let us find the separation points of the three parts of the signal,
denoted by $t_{\rm th}^{\pm}$. Assume that $|\delta|$ is very small such that
the two first-order solitons are far from the middle part. Then, 
the middle part converges to a multi-soliton with $\{(\lambda_k,b_k)\}_{k=1}^{n-1}$.
The separation point $t_{\rm th}^{+}$ is approximately equal to the intersection point of the right tail of the middle multi-soliton, given in \eqref{eq:tail}, and the left tail of the first-order soliton on the right, given in \eqref{eq:qtilde_tails}. Similarly, we can approximate the separation point $t_{\rm th}^{-}$. Accordingly, we obtain
\begin{align}\label{eq:t_th}
& t_{\mathrm{th}}^{\pm} \approx \pm \frac{ 0.5 \ln \left(|b_{n-1}|^{\pm 1}\frac{\sigma_{n-1}}{\sigma_n} \right) + \sigma_n t_\delta^{\pm} + \sigma_{n-1} t_0 }{\sigma_n+\sigma_{n-1}} 
\\
& \mathrm{where} \qquad
t_0=\frac{1}{2\sigma_{n-1}} \ln \left(\left| \prod_{k=1}^{n-2} \frac{\lambda_{n-1}-\lambda_k^*}{\lambda_{n-1}-\lambda_k} \right| \right)
\label{eq:t0}
\end{align}
and $\sigma_{n-1}=\text{Im}\{\lambda_{n-1}\}$.

Now we find a condition on $|\delta|$. After addition of $\hat{\lambda}_n$, the SER algorithm truncates $\tilde{q}(t)$ to the interval $[T_-^{(n-1)},T_+^{(n-1)}]$. A necessary condition for the truncated signal to contain only the middle part of the signal is $[T_-^{(n-1)},T_+^{(n-1)}]\subset[t_{\mathrm{th}}^{-},t_{\mathrm{th}}^{+}]$.
To satisfy this condition, $|\tilde{q}(t_{\mathrm{th}}^{\pm})|<2\sigma_{n-1}\sqrt{\varepsilon}$. From \eqref{eq:tail} and \eqref{eq:t_th}, we find,
\begin{align}
\left|\frac{\delta}{\sigma_n}\right| \leq\max\frac{|\alpha^\pm|\sqrt{\varepsilon}\sigma_{n-1}}{\sigma_n} \left(\frac{\sqrt{\varepsilon}}{2|b_{n-1}|^{\pm 1}}\right)^{\frac{\sigma_n}{\sigma_{n-1}}}e^{2\sigma_n(t_s-t_0)}
\end{align}
where $t_s$ and $t_0$ are given in \eqref{eq:ts} and \eqref{eq:t0}.

An interesting advantage of the SER algorithm is that it allows to \emph{detect} whether an eigenvalue has been estimated precisely enough. By removing the eigenvalue $\lambda_n$, the pulse energy must decrease by $4\text{Im}(\lambda_n)$. This can be verified by checking the pulse energy after the truncation. Such a capability is missing in other usual NFT algorithms~\cite{yousefi2014information,aref2016control,Hari2016,Vasylchenkova2018,garcia2019statistics,Chimmalgi2018,chimmalgi2019fast,Medvedev20,vaibhav2019efficient,Mullyadzhanov2019}.

\section{Simulation Results}\label{sec:Sim_eval}

We compare the SER algorithm to the common NFT methodology 
that finds all eigenvalues and their 
spectral coefficients separately without pulse modification.
We compare them in terms of computational complexity and estimation accuracy 
for some exemplary multi-soliton pulses. Note that the absolute accuracy and/or complexity depends on the specific NFT method that is used. However, the SER algorithm can be applied with any underlying NFT method. Thus, although we evaluate all results based on the method in Sec.\ref{sec:NFT}, the advantage of the SER algorithm is quite general.

\begin{figure}[tb]
\includegraphics{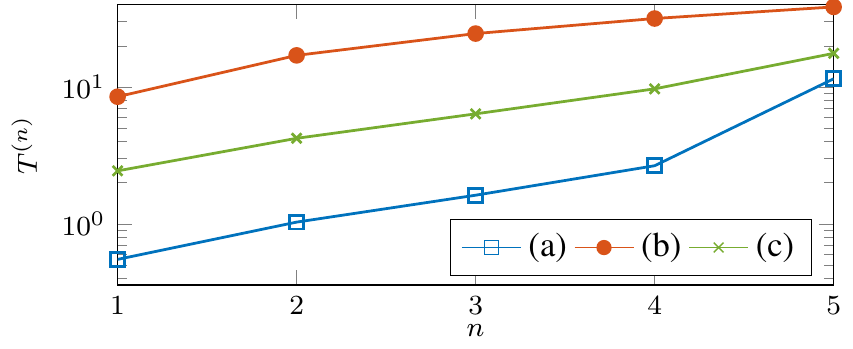}
%
%
%
%
%
%
%
%
%
%
%

\caption{Pulse duration $T^{(n)}$ for $n$ remaining eigenvalues after each iteration of the SER algorithm. Three multi-soliton pulses are considered with  
(a) $\Lambda_a^{(5)}=\{9j;7j;5j;3j;0.5j\}$, (b) $\Lambda_b^{(5)}=\{0.58j;0.56j;0.54j;0.52j;0.5j\}$ and (c) $\Lambda_c^{(5)}=\{2.5j;2j;1.5j;1j;0.5j\}$.}
\label{fig:5Soliton_Tmax_vs_N}
\end{figure}

\subsection{Pulse Duration and Complexity Reduction}

When the ZSS is solved numerically, the computational complexity scales linearly in terms of the pulse samples $M$.
The classical approaches solve the ZSS for all $N$ eigenvalues based on the same given pulse. Although all $N$ eigenvalues can be computed in parallel,
the total computational complexity is $C N M$ for some constant $C$ operation cost per sample\footnote{We assumed the eigenvalues are given. 
To estimate the eigenvalues by a Newton search, the total complexity is multiplied by 
the iteration number of the Newton method. We neglect this multiplication
factor for simplicity as it is almost the same for the SER algorithm and the other methods.}. 

The SER algorithm truncates the pulse support after removing each eigenvalue. Recall that $T^{(n)}=T^{(n)}_+-T^{(n)}_-$ is the pulse duration when the pulse has still $n$ eigenvalues. Let $T^{(N)}$ denote the initial pulse duration of a pulse with $N$ eigenvalues. Then the computational complexity is $C M\frac{T^{(n)}}{T^{(N)}}$ if the same NFT algorithm for computing the spectral 
coefficient of $\lambda_n$ is applied. The total complexity of the SER algorithm is then $\alpha_N C N M$ with $\alpha_N=\frac{\sum_{n=1}^NT^{(n)}}{N T^{(N)}}$. 

The SER algorithm has a factor $\alpha_N$ smaller complexity than a classical approach. The complexity gain, however, depends on the nonlinear spectrum of the pulse.
Fig.~\ref{fig:5Soliton_Tmax_vs_N} shows how the pulse duration $T^{(n)}$ decreases after each SER iteration for three exemplary multi-solitons with $5$ eigenvalues.
Three sets of eigenvalues are chosen: $(a)$ $\Lambda_a^{(5)}=\{9j;7j;5j;3j;0.5j\}$, $(b)$  $\Lambda_b^{(5)}=\{0.58j;0.56j;0.54j;0.52j;0.5j\}$, and $(c)$ $\Lambda_c^{(5)}=\{2.5j;2j;1.5j;1j;0.5j\}$. For all three pulses, we let $b_k=(-1)^k$ and $\varepsilon=2\cdot 10^{-4}$.
For the chosen soliton examples, the complexity reduces by $\alpha_N$ $(a)$ $\approx 0.3$, $(b)$ $\approx 0.62$ and $(c)$ $\approx 0.46$, respectively. Note that from \eqref{eq:tail}, the pulse duration $T^{(n)}$ can be well approximated as
\begin{equation}
    T^{(n)}\approx 2t_s + \frac{1}{2\sigma_n}\ln\left(\frac{4}{\varepsilon}\right)
\end{equation}
for sufficiently small $\varepsilon$. We see that, up to the first-order approximation, 
$T^{(n)}$ is independent of the spectral phases $\varphi_k$.

The FC method was not considered in the complexity analysis as its complexity is independent of the SER algorithm. Note that, before their refinement, only a coarse estimate of the eigenvalues is needed to determine their removal order. The FC can provide such a coarse estimate with only a few samples~\cite{garcia2019statistics}. Thus, although it has cubic complexity with respect to the number of samples, the SER's total complexity is not dominated by the FC method. If predefined eigenvalues are used for modulation, FC is not needed at all.

\subsection{Estimation Accuracy}

\begin{figure}
\includegraphics{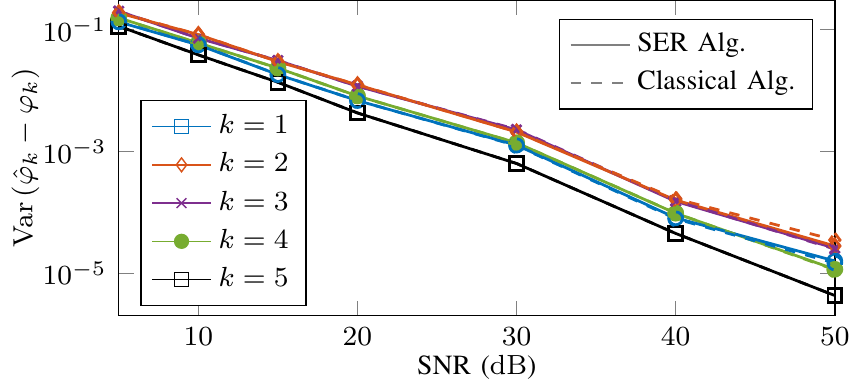}

\caption{Accuracy of the SER algorithm in comparison to the classical NFT algorithm of Sec.~\ref{sec:NFT} for the detection of the spectral amplitudes $b(\lambda_k)=\exp(j\phi_k)$.
The variance of the phase estimation error $\mathrm{Var}\left(\hat{\varphi}_k - \varphi_k\right)$
in terms of SNR for a multi-soliton pulse with $\Lambda_c^{(5)}=\{2.5j;2j;1.5j;1j;0.5j\}$ being perturbed by an additive white Gaussian noise is shown.
}
\label{fig:5Soliton_variance_vs_snr}
\end{figure}

We compare the precision of the SER algorithm and the common NFT methodology in finding the spectral coefficients of a given pulse. To have a fair comparison, the SER algorithm should use the same NFT algorithm to find $\hat{\boldsymbol{\vartheta}}^{(n)}(\hat{\lambda}_n;t)$.
Here, we use the NFT algorithm of Sec.~\ref{sec:NFT}.
We consider the eigenvalue set $\Lambda_c^{(5)}$ with $b_k=\exp(j\varphi_k)$. 
We generate $5000$ multi-soliton pulses $q^{(5)}(t)$ with randomly chosen $\varphi_k$ values.
Let us define the bandwidth $B$ of a pulse in (linear) Fourier domain 
as the frequency support containing $99.99\%$ of the total energy.
All pulses are sampled with the same sampling rate of $f_s=4 \max_{\varphi_k}{B}$. 
We further add a white Gaussian noise to the pulses according to a signal-to-noise ratio (SNR) 
defined within the bandwidth $B_\mathrm{max}$. Then, we compute the spectral coefficients $\hat{b}(\hat{\lambda}_k)$ of the noisy pulses using both algorithms. Define $\hat{\varphi}_k=\arg \{\hat{b}(\hat{\lambda}_k)\}$.
Fig.~\ref{fig:5Soliton_variance_vs_snr} shows the variance of the phase error, $\mathrm{Var}\left(\hat{\varphi}_k - \varphi_k\right)$, in terms of SNR for both algorithms. We observe that both algorithms have quite the same precision while the SER algorithm has about 
half the complexity.

\begin{figure}
\includegraphics{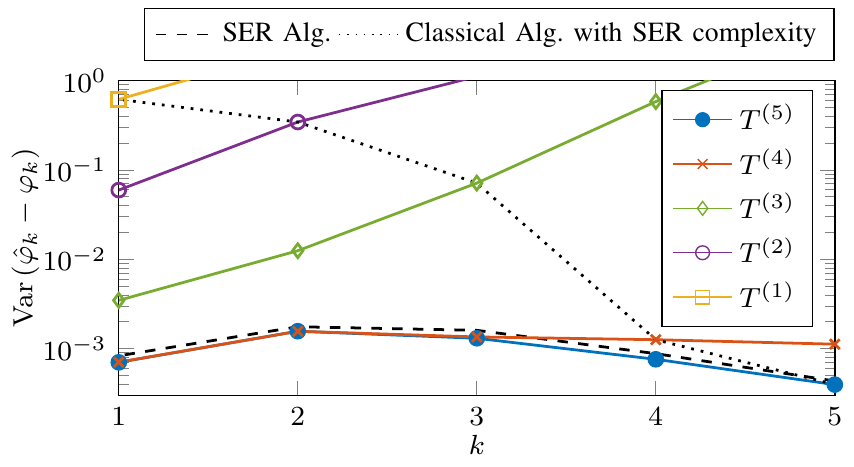}

\caption{Effect of a pulse truncation $T^{(n)}$ on the accuracy of a classical NFT algorithm to detect $b(\lambda_k)=\exp(j\phi_k)$. The variance of the phase estimation error $\mathrm{Var}\left(\hat{\varphi}_k - \varphi_k\right)$ is shown for a randomly phase modulated multi-soliton pulse with $\Lambda_c^{(5)}=\{2.5j;2j;1.5j;1j;0.5j\}$ with an additive white Gaussian noise perturbation at SNR$=30\mathrm{dB}$. The dashed line shows the result when using the SER algorithm. The dotted line is the accuracy of the classical NFT algorithm when applying the same truncation value $T^{(n)}$ as in the SER algorithm before finding the respective $\hat{b}(\hat{\lambda}_n)$.
}
\label{fig:5Soliton_variance_vs_truncation}
\end{figure}

The SER algorithm reduces the complexity by pulse truncation.
We now compare the performance of the SER and the classical algorithm when both algorithms have the same total number of samples (and thus the same complexity). 
One interesting scenario is to investigate how the performance of the classical NFT algorithm is sensitive to the pulse truncation. The logic behind this scenario is to see 
how the contribution of an eigenvalue is distributed over the support of an initial pulse.
We choose the pulse truncations according to the SER algorithm: $T^{(5)},T^{(4)},T^{(3)},T^{(2)},T^{(1)}$ as given in Fig.~\ref{fig:5Soliton_Tmax_vs_N}. The pulses are generated and sampled as explained before at SNR$=30\mathrm{dB}$.
We compute the spectral coefficients using the classical NFT methodology when applying the different truncations $T^{(n)}$. Fig.~\ref{fig:5Soliton_variance_vs_truncation} shows the respective $\mathrm{Var}\left(\hat{\varphi}_k - \varphi_k\right)$ for all eigenvalues with different truncation.  
The dashed line indicates the performance of the SER algorithm.
The dotted line is the performance of the classical NFT algorithm when it applies the same truncation value $T^{(n)}$ as in the SER algorithm before finding each $\hat{b}(\hat{\lambda}_n)$. This way, it uses the same number of samples as the SER algorithm but its performance degrades significantly.
In fact, any pulse truncation causes a drastic growth in the estimation error without applying the SER pulse modification. On the other hand, by applying the pulse modification of SER algorithm, the nonlinear spectral contents will be condensed in a shorter time support.

\section{Conclusion}\label{sec:concl}

We proposed the successive eigenvalue removal algorithm to compute the discrete spectrum.
The algorithm computes the spectral coefficients successively and carefully
removes the previously computed eigenvalue and its spectral coefficient from the spectrum.
The algorithm peels off the eigenvalues of a pulse one by one to simplify the computation of the remaining discrete spectrum. As an application, we showed a considerable complexity gain of the SER algorithm compared to the conventional methodology without pulse modification in finding the spectral coefficients without sacrificing accuracy.


%
%



%
\bibliographystyle{IEEEtran}
\bibliography{references_nft}

\end{document}